%% file: main.tex
\documentclass[runningheads]{llncs}
\input{commands}

\title{Snap-and-Chat Protocols: System Aspects}
\author{%
Joachim Neu%
\and
Ertem Nusret Tas%
\and
David Tse%
}
\authorrunning{J. Neu \and E. N. Tas \and D. Tse}

\institute{%
\email{\{jneu,nusret,dntse\}@stanford.edu}%
}

\begin{document}

\maketitle
\begingroup
\renewcommand\thefootnote{}%
\footnotetext{JN and ENT contributed equally and are listed alphabetically. Contact author: DT.}
\endgroup

\begin{abstract}
The availability-finality dilemma \cite{lewispye2020resource} says that blockchain protocols cannot be both available under dynamic participation and safe under network partition. \Sac protocols have recently been proposed as a resolution to this dilemma \cite{ebbandflow}. A \sac protocol produces an always available ledger containing a finalized prefix ledger which is always safe and catches up with the available ledger whenever network conditions permit. In contrast to existing handcrafted finality gadget based designs like Ethereum 2.0's consensus protocol Gasper \cite{buterin2020combining}, \sac protocols are constructed as a black-box composition of off-the-shelf BFT and longest chain protocols.
In this paper, we consider system aspects of \sac protocols and show how they can provide two important features: 1) accountability, 2) support of light clients.
Through this investigation, a deeper understanding of the strengths and challenges of \sac protocols is gained.
\end{abstract}

\input{introduction}

\input{snap_and_chat_protocols}

\input{accountability}

\input{light_clients}

\section*{Acknowledgment}
We thank Sreeram Kannan for fruitful discussions.
JN is supported by the Reed-Hodgson Stanford Graduate Fellowship.
ENT is supported by the Stanford Center for Blockchain Research.

\bibliographystyle{splncs04}
\bibliography{references}

\appendix

\input{appendix_accountability_hotstuff}

\input{appendix_lightclient}

\end{document}

%% file: commands.tex
\usepackage{algorithmicx}
\usepackage{algpseudocode}
\usepackage{algorithm}

\PassOptionsToPackage{hyphens}{url}
\usepackage{url}
\usepackage{hyperref}
\hypersetup{breaklinks=true}

\usepackage[utf8]{inputenc}
\usepackage[T1]{fontenc}

\usepackage{lmodern}

\usepackage{amssymb,amsfonts,amsmath,amsthm}
\usepackage{cite}   %

\usepackage{graphicx}
\usepackage{xcolor}

\usepackage[caption=false,font=footnotesize]{subfig}
\usepackage{float}

\usepackage{IEEEtrantools}
\allowdisplaybreaks

\usepackage[shortlabels,inline]{enumitem}

\usepackage{xstring}

\usepackage{datetime2}

\usepackage{siunitx}

\usepackage{varwidth}

\usepackage{tikz}
\usetikzlibrary{calc}
\usetikzlibrary{arrows}
\usetikzlibrary{arrows.meta}
\usetikzlibrary{patterns}
\usetikzlibrary{positioning}
\usetikzlibrary{decorations.pathreplacing}
\usetikzlibrary{shapes.misc}
\usetikzlibrary{spy}

\usepackage{pgfplots}
\pgfplotsset{compat=1.14}
\usepgfplotslibrary{fillbetween}

\usepackage{algorithmicx}
\usepackage{algpseudocode}
\usepackage{algorithm}

\usepackage{xspace}
\usepackage{ifthen}

\newcommand{\sac}{snap-and-chat\xspace}
\newcommand{\Sac}{Snap-and-chat\xspace}

\newcommand{\eaf}{ebb-and-flow\xspace}

\newcommand{\blocklc}[0]{LC block\xspace}
\newcommand{\blockslc}[0]{LC blocks\xspace}
\newcommand{\blockbft}[0]{BFT block\xspace}
\newcommand{\blocksbft}[0]{BFT blocks\xspace}

\newcommand{\LOGda}[2]{%
    \ifthenelse{\equal{#1}{}}{%
        \ensuremath{\mathsf{LOG}_{\mathrm{da}}^{#2}}%
    }{%
        \ensuremath{\mathsf{LOG}_{\mathrm{da},#1}^{#2}}%
    }%
}
\newcommand{\LOGfin}[2]{%
    \ifthenelse{\equal{#1}{}}{%
        \ensuremath{\mathsf{LOG}_{\mathrm{fin}}^{#2}}%
    }{%
        \ensuremath{\mathsf{LOG}_{\mathrm{fin},#1}^{#2}}%
    }%
}

\newcommand{\LOGlc}[2]{%
    \ifthenelse{\equal{#1}{}}{%
        \ensuremath{\mathsf{LOG}_{\mathrm{lc}}^{#2}}%
    }{%
        \ensuremath{\mathsf{LOG}_{\mathrm{lc},#1}^{#2}}%
    }%
}
\newcommand{\LOGbft}[2]{%
    \ifthenelse{\equal{#1}{}}{%
        \ensuremath{\mathsf{LOG}_{\mathrm{bft}}^{#2}}%
    }{%
        \ensuremath{\mathsf{LOG}_{\mathrm{bft},#1}^{#2}}%
    }%
}
\newcommand{\LOG}[2]{%
    \ensuremath{\mathsf{LOG}_{#1}^{#2}}
}

\newcommand{\PIlc}[0]{\ensuremath{\Pi_{\mathrm{lc}}}}
\newcommand{\PIbft}[0]{\ensuremath{\Pi_{\mathrm{bft}}}}
\newcommand{\PIsac}[0]{\ensuremath{\Pi_{\mathrm{sac}}}}

\newcommand{\PIanyexample}[0]{\ensuremath{\Pi_{\mathrm{sac}}}}
\newcommand{\PItheexample}[2]{\ensuremath{\PIanyexample}}

\newcommand{\Txs}[2]{\ensuremath{\mathsf{txs}_{#1}^{#2}}}

\newcommand{\ie}[0]{\emph{i.e.}\xspace}
\newcommand{\eg}[0]{\emph{e.g.}\xspace}

\newcommand{\GST}[0]{\ensuremath{\mathsf{GST}}}
\newcommand{\GOT}[0]{\ensuremath{\mathsf{GAT}}}

\newcommand{\supersanitize}[0]{\ensuremath{\operatorname{Sanitize}}}
\newcommand{\merkleize}[0]{\ensuremath{\operatorname{Merkleize}}}
\newcommand{\AcceptMerkleProofWrt}[0]{\ensuremath{\operatorname{AcceptOpeningsWrt}}}

\newcommand{\tx}[0]{\ensuremath{\mathrm{tx}}}

\theoremstyle{plain}
\newtheorem*{theorem*}{Theorem}

\definecolor{myParula01Blue}{RGB}{0,114,189}
\definecolor{myParula02Orange}{RGB}{217,83,25}
\definecolor{myParula03Yellow}{RGB}{237,177,32}
\definecolor{myParula04Purple}{RGB}{126,47,142}
\definecolor{myParula05Green}{RGB}{119,172,48}
\definecolor{myParula06LightBlue}{RGB}{77,190,238}
\definecolor{myParula07Red}{RGB}{162,20,47}

\tikzset{myparula11/.style={color=myParula01Blue,solid,mark=+,mark options={solid}}}
\tikzset{myparula12/.style={color=myParula01Blue,densely dashed,mark=x,mark options={solid}}}
\tikzset{myparula13/.style={color=myParula01Blue,densely dotted,mark=o,mark options={solid}}}
\tikzset{myparula14/.style={color=myParula01Blue,dashdotted,mark=triangle,mark options={solid}}}
\tikzset{myparula15/.style={color=myParula01Blue,dashdotdotted,mark=square,mark options={solid}}}

\tikzset{myparula21/.style={color=myParula02Orange,solid,mark=+,mark options={solid}}}
\tikzset{myparula22/.style={color=myParula02Orange,densely dashed,mark=x,mark options={solid}}}
\tikzset{myparula23/.style={color=myParula02Orange,densely dotted,mark=o,mark options={solid}}}
\tikzset{myparula24/.style={color=myParula02Orange,dashdotted,mark=triangle,mark options={solid}}}
\tikzset{myparula25/.style={color=myParula02Orange,dashdotdotted,mark=square,mark options={solid}}}

\tikzset{myparula31/.style={color=myParula03Yellow,solid,mark=+,mark options={solid}}}
\tikzset{myparula32/.style={color=myParula03Yellow,densely dashed,mark=x,mark options={solid}}}
\tikzset{myparula33/.style={color=myParula03Yellow,densely dotted,mark=o,mark options={solid}}}
\tikzset{myparula34/.style={color=myParula03Yellow,dashdotted,mark=triangle,mark options={solid}}}
\tikzset{myparula35/.style={color=myParula03Yellow,dashdotdotted,mark=square,mark options={solid}}}

\tikzset{myparula41/.style={color=myParula04Purple,solid,mark=+,mark options={solid}}}
\tikzset{myparula42/.style={color=myParula04Purple,densely dashed,mark=x,mark options={solid}}}
\tikzset{myparula43/.style={color=myParula04Purple,densely dotted,mark=o,mark options={solid}}}
\tikzset{myparula44/.style={color=myParula04Purple,dashdotted,mark=triangle,mark options={solid}}}
\tikzset{myparula45/.style={color=myParula04Purple,dashdotdotted,mark=square,mark options={solid}}}

\tikzset{myparula51/.style={color=myParula05Green,solid,mark=+,mark options={solid}}}
\tikzset{myparula52/.style={color=myParula05Green,densely dashed,mark=x,mark options={solid}}}
\tikzset{myparula53/.style={color=myParula05Green,densely dotted,mark=o,mark options={solid}}}
\tikzset{myparula54/.style={color=myParula05Green,dashdotted,mark=triangle,mark options={solid}}}
\tikzset{myparula55/.style={color=myParula05Green,dashdotdotted,mark=square,mark options={solid}}}

\tikzset{myparula61/.style={color=myParula06LightBlue,solid,mark=+,mark options={solid}}}
\tikzset{myparula62/.style={color=myParula06LightBlue,densely dashed,mark=x,mark options={solid}}}
\tikzset{myparula63/.style={color=myParula06LightBlue,densely dotted,mark=o,mark options={solid}}}
\tikzset{myparula64/.style={color=myParula06LightBlue,dashdotted,mark=triangle,mark options={solid}}}
\tikzset{myparula65/.style={color=myParula06LightBlue,dashdotdotted,mark=square,mark options={solid}}}

\tikzset{myparula71/.style={color=myParula07Red,solid,mark=+,mark options={solid}}}
\tikzset{myparula72/.style={color=myParula07Red,densely dashed,mark=x,mark options={solid}}}
\tikzset{myparula73/.style={color=myParula07Red,densely dotted,mark=o,mark options={solid}}}
\tikzset{myparula74/.style={color=myParula07Red,dashdotted,mark=triangle,mark options={solid}}}
\tikzset{myparula75/.style={color=myParula07Red,dashdotdotted,mark=square,mark options={solid}}}

\pgfplotsset{
    mysimpleplot/.style = {
        every axis plot/.prefix style={thick},
        width=\linewidth,
        height=0.75\linewidth,
        title style={font=\footnotesize,align=center},
        legend cell align=left,
        legend style={font=\footnotesize},
        legend columns=3,
        legend style={
            at={(0.5,1)},
            yshift=0.3em,
            anchor=south,
            draw=none,
            /tikz/every even column/.append style={
                column sep=0.3em
            },
            cells={
                align=left
            }
        },
        grid=both,
        minor tick num=4,
        major grid style={solid,draw=gray!50},
        minor grid style={densely dotted,draw=gray!50},
        label style={font=\footnotesize,align=center},
        tick label style={font=\footnotesize},
    },
}

\usepackage{pifont}%
\newcommand{\cmark}{\ding{52}}%
\newcommand{\xmark}{\ding{56}}%

\usepackage{marvosym}

\usepackage{centernot}
\newcommand{\concatminus}[0]{\ensuremath{\setminus}}

%% file: introduction.tex
\section{Introduction}
\label{sec:introduction}

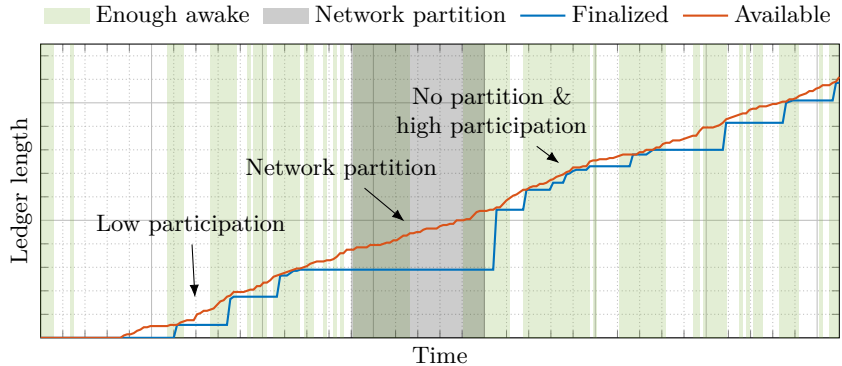
\begin{figure}[tb]
    \centering
    \begin{tikzpicture}
        \footnotesize
        \begin{scope}[spy using outlines={rectangle,width=5cm,height=2.5cm,gray,magnification=3,connect spies}]
            \begin{axis}[
                mysimpleplot,
                name=plot1,
                xlabel={Time},
                ylabel={Ledger length},
                legend columns=4,
                xmin=0, xmax=3600,
                ymin=0, ymax=250,
                height=0.45\linewidth,
                xmajorticks=false,
                ymajorticks=false,
            ]
            
                \def\DATAPREFIX{./figures/simulation-04-intro}

                \foreach \tStart/\tMidway/\tStop in {0/30.0/60, 135/142.5/150, 570/607.5/645, 765/825.0/885, 930/937.5/945, 960/990.0/1020, 1050/1110.0/1170, 1185/1207.5/1230, 1275/1282.5/1290, 1320/1327.5/1335, 1350/1357.5/1365, 1410/1537.5/1665, 1905/2010.0/2115, 2175/2325.0/2475, 2490/2497.5/2505, 2610/2715.0/2820, 2940/2955.0/2970, 2985/3037.5/3090, 3150/3157.5/3165, 3180/3187.5/3195, 3210/3232.5/3255, 3330/3375.0/3420, 3510/3517.5/3525, 3555/3577.5/3600} {
                    \edef\temp{\noexpand\draw [fill=myParula05Green,fill opacity=0.2,draw=none] (axis cs:\tStart,-100) rectangle (axis cs:\tStop,1000);}
                    \temp
                }
                
                \addlegendimage{area legend,draw opacity=0,fill=myParula05Green,fill opacity=0.2,draw=none};
                \addlegendentry{Enough awake};

                \draw [fill=black,fill opacity=0.2,draw=none] (axis cs:1400,-100) rectangle (axis cs:2000,1000);
    
                \addlegendimage{area legend,draw opacity=0,fill=black,fill opacity=0.2,draw=none};
                \addlegendentry{Network partition};

                \addplot [myparula11,mark=none] table [x=t,y=l_Lp]
                {\DATAPREFIX/sim-04-phases.dat};
                \label{leg:simulations-intro-Lp}
                \addlegendentry{Finalized}; %
                
                \addplot [myparula21,mark=none] table [x=t,y=l_Lda]
                {\DATAPREFIX/sim-04-phases.dat};
                \label{leg:simulations-intro-Lda}
                \addlegendentry{Available}; %

                \coordinate (pointallgood) at (axis cs:2400,140);
                \coordinate (pointpartition) at (axis cs:1700,90);
                \coordinate (pointlowparticipation) at (axis cs:700,20);
                
            \end{axis}

            \draw node at (6,3) [align=center] {No partition \&\\ high participation} edge [-Latex,shorten >=0.7em] (pointallgood);
            \draw node at (4,2.25) [align=center] {Network partition} edge [-Latex,shorten >=0.7em] (pointpartition);
            \draw node at (2,1.5) [align=center] {Low participation} edge [-Latex,shorten >=0.7em] (pointlowparticipation);
            
        \end{scope}
    \end{tikzpicture}
    \vspace{-0.25cm}
    \caption[]{%
        Simulated run of a \sac protocol under dynamic participation and network partition. Lengths of the two ledgers are plotted over time. During network partition or when few nodes are awake, the finalized ledger falls behind the available ledger, but catches up after the network heals or when a sufficient number of nodes wake up.
        Modified from Figure 2 in \cite{ebbandflow}.%
    }
    \label{fig:intro_sim}
\end{figure}

Due to a variant \cite{lewispye2020resource} of the infamous
CAP theorem \cite{cap}, there cannot be
a secure state machine replication (SMR) consensus protocol
with transactions as input and a ledger as output
in an environment that exhibits both
dynamic participation of validators
and network partitions.
Due to this dilemma, protocol designers
had to decide whether to favor liveness or safety.
To obviate a system-wide determination
and leave the choice to the end-user,
recent constructions output
not a single ledger, %
but two ledgers,
a dynamically \emph{available} (full) \emph{ledger} %
in conjunction with a \emph{finalized} (prefix) \emph{ledger}. %
When conditions permit, both ledgers are secure.
During periods of low participation or network partitions
(\ie, when the CAP theorem is an active constraint),
the available ledger
remains live
but might suffer from
inconsistencies,
while the finalized ledger
remains safe
but might stall
(Figure~\ref{fig:intro_sim}).
Once the environment has returned to favorable conditions,
the available ledgers
across nodes
are reconciled to a single `common account of history'
and the finalized ledger
catches up with
the available ledger.
Users choose whether to favor liveness (adopt the available ledger) %
or safety (adopt the finalized ledger). %
Eventually, all users will
agree on a single history but in the process might
temporarily lose safety or liveness, if network conditions
do not allow otherwise.
The above mechanics and properties
have recently been formalized as
an
\emph{\eaf property} in \cite{ebbandflow}.

Gasper \cite{buterin2020combining},
the current candidate %
for Ethereum 2.0's beacon chain,
combines the finality gadget Casper FFG \cite{buterin2017casper}
with the latest message driven (LMD)
GHOST
fork choice rule,
and aims to be a secure \eaf protocol.
However, \cite{ebbandflow,ethresearch-gasper-attack}
demonstrated a bouncing-type liveness attack on Gasper
in the synchronous network model.
The attack causes Gasper to lose liveness of
the finalized ledger
and safety of
the available ledger
indefinitely,
raising concerns about
its
suitability
for the Ethereum 2.0 beacon chain.
In the same work,
\emph{\sac protocols} have been proposed
as a provably secure alternative,
readily constructed via an almost-black-box composition of
an off-the-shelf
permissioned
dynamically available
protocol
such as
\cite{sleepy,david2018ouroboros}
and an off-the-shelf partially synchronous Byzantine fault tolerant (BFT) consensus protocol
such as \cite{yin2018hotstuff,streamlet,pbft}.

However, besides security,
many features required for
an Internet-scale open-participation
consensus infrastructure such as Ethereum 2.0
have been left to be desired in \cite{ebbandflow}.
For instance,
aspects of cryptoeconomic security and
incentive compatibility have been neglected in \cite{ebbandflow}.
Furthermore,
since \sac protocols aim to be an alternative
to Gasper as Ethereum 2.0's beacon chain protocol,
support for light clients is imperative
but missing in \cite{ebbandflow}.

\paragraph{Contributions}

We contribute solutions to the aforementioned shortcomings
of \sac protocols.
In particular, we show:
\begin{enumerate}
    \item
        \Sac protocols, when constructed from
        typical BFT protocols
        following the propose-and-vote paradigm,
        such as
        HotStuff \cite{yin2018hotstuff} or Streamlet \cite{streamlet},
        provide \emph{accountable safety}.
        That is,
        a safety violation implies that at least a third of
        validators have provably violated the protocol's
        so called \emph{slashing conditions}.
        As a
        punitive and deterrent, those validators'
        stake can be slashed,
        and since stake carries value,
        this attaches a price tag to safety violations
        and provides a notion of cryptoeconomic security.
    
    \item
        Many users of, \eg, a blockchain-based cryptocurrency, are only
        interested in the outcome of a small subset of the transactions.
        These users should not have to download all of the system's data.
        Instead, it should suffice if they follow only the block headers
        (\emph{light clients}) with respect to which they can later
        convince themselves (using auxiliary information furnished
        by an untrusted
        \emph{full node} that maintains the whole system state)
        of the outcome of the transactions of interest.
        Standard approaches to supporting light clients fall short
        for \sac protocols due to particularities
        of the ledger extraction.
        We show how to tweak the standard approaches to apply them
        to \sac protocols to obtain light client support.
\end{enumerate}

How we provide these features
turns the spotlight on strengths and challenges
of \sac protocols.
Due to the modular construction of \sac protocols,
some features of the constituent protocols
are readily inherited by the \sac protocol
-- a strength of the approach
exemplified by our treatment of accountability.
On the flip side,
in contrast to finality gadget (FG) based constructions,
the more involved ledger extraction procedure of \sac protocols
brings new challenges as it
breaks, \eg, standard approaches
to light client support.
We showcase how to overcome this challenge
with a twist to the current techniques.

\paragraph{Outline}

In Section~\ref{sec:protocols},
we review the construction of \sac protocols
and the \eaf security property they provide, so as to make the present paper self-contained.
We also contrast \sac protocols with finality-gadget-based constructions
and highlight resulting strengths and challenges.
We show in Section~\ref{sec:accountability}
how %
accountable safety can be readily inherited
from the BFT sub-protocol. %
To serve as a drop-in replacement, \eg, for Gasper as
Ethereum 2.0's beacon chain protocol,
it is important
to support
light clients,
which we enable with a twist to standard techniques, shown %
in Section~\ref{sec:light-clients}.

%% file: snap_and_chat_protocols.tex
\section{Snap-and-Chat Protocols}
\label{sec:protocols}

In this section, we briefly recapitulate the construction of \sac protocols
and the \eaf security property they provide,
as detailed in \cite{ebbandflow}.
We also compare \sac protocols with finality gadget based designs
such as \cite{buterin2017casper,buterin2020combining}
and highlight strengths and challenges of \sac protocols.

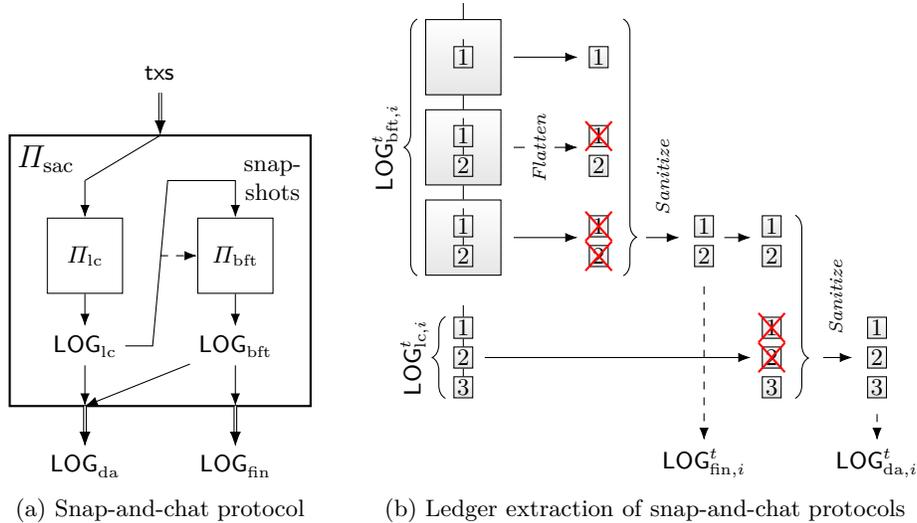
\begin{figure} 
    \centering
    \subfloat[Snap-and-chat protocol\label{fig:protocol-overview-eaf}]{%
    \begin{tikzpicture}[
            x=0.2cm,
            y=0.4cm,
        ]
        \small
        
        \node (box_lc) at (-5,0) [draw,minimum width=1cm,minimum height=1cm,align=center] {$\PIlc$};
        \node (box_bft) at (+5,0) [draw,minimum width=1cm,minimum height=1cm,align=center] {$\PIbft$};
        
        \node (chain_lc) at (-5,-3) [align=center] {$\LOGlc{}{}$};
        \node (chain_bft) at (5,-3) [align=center] {$\LOGbft{}{}$};

        \coordinate (txs) at (-5,2.5);

        \coordinate (ledger_da) at (-5,-5);
        \coordinate (ledger_fin) at (5,-5);
        
        \draw[-Latex] (txs) -- (box_lc);
        \draw[-Latex] (box_lc) -- (chain_lc);
        \draw[-Latex] (box_bft) -- (chain_bft);
        
        \draw[-Latex] (chain_lc) -- (-0.5,-3) -- (0.5,2.5) -- (5,2.5) -- (box_bft)
            node [pos=0,right,align=center,font=\footnotesize] {snap-\\shots};
        
        \draw[-Latex,dashed] (0,0) -- (box_bft);

        \draw[-Latex] (chain_lc) -- (ledger_da);
        \draw[-Latex] (chain_bft) -- (ledger_da);
        \draw[-Latex] (chain_bft) -- (ledger_fin);

        \node (outerbox) at (0,-0.5) [draw,minimum height=3.6cm,minimum width=4cm,thick] {};
        
        \coordinate (INTERFACE_txs_connector) at (0,4);
        \node (INTERFACE_txs) at (0,6) [align=center] {$\Txs{}{}$};
        
        \draw [-Latex,double] (INTERFACE_txs) -- (INTERFACE_txs_connector);
        \draw (INTERFACE_txs_connector) -- (txs);
        
        \coordinate (INTERFACE_LOGda_connector) at (-5,-5);
        \coordinate (INTERFACE_LOGfin_connector) at (5,-5);
        \node (INTERFACE_LOGda) at (-5,-7) {$\LOGda{}{}$};
        \node (INTERFACE_LOGfin) at (5,-7) {$\LOGfin{}{}$};
        
        \draw [-Latex,double] (INTERFACE_LOGda_connector) -- (INTERFACE_LOGda);
        \draw [-Latex,double] (INTERFACE_LOGfin_connector) -- (INTERFACE_LOGfin);

        \node at (outerbox.north west) [anchor=north west,yshift=-1pt] {\large $\PIanyexample$};

    \end{tikzpicture}%
    }%
    \hfill%
    \subfloat[Ledger extraction of \sac protocols\label{fig:ledger-extraction}]{%
        \begin{tikzpicture}[
            x=1.8cm,
            y=1.2cm,
        ]
        \footnotesize
        
        \tikzset{
            _block/.style = {
                draw,
                shade,
                top color=white,
                bottom color=black!10,
            },
            largeblock/.style = {
                _block,
                minimum width=1cm,
                minimum height=1cm,
                draw,
            },
            largelink/.style = {
                draw,
            },
            smallblock/.style = {
                _block,
                minimum width=0.25cm,
                minimum height=0.28cm,
                draw,
                inner sep=0,
                text=black,
            },
            smalllink/.style = {
                draw,
            },
        }

        \coordinate (blgen) at (0,0.60);
        \node (bl1) at (0,0) [largeblock] {};
        \node (bl2) at (0,-1) [largeblock] {};
        \node (bl3) at (0,-2) [largeblock] {};
        
        \draw [largelink] (bl3) -- (bl2);
        \draw [largelink] (bl2) -- (bl1);
        \draw [largelink] (bl1) -- (blgen);

        \coordinate (bsgen) at (0,-2.75);
        \node (bs1) at (0,-3) [smallblock] {1};
        \node (bs2) at (0,-3.33) [smallblock] {2};
        \node (bs3) at (0,-3.66) [smallblock] {3};
        
        \draw [largelink] (bs3) -- (bs2);
        \draw [largelink] (bs2) -- (bs1);
        \draw [largelink] (bs1) -- (bsgen);
        
        \draw [decorate,decoration={brace,amplitude=5pt}]
        ([xshift=-3pt]bl3.south west) -- ([xshift=-3pt]bl1.north west)
        node [midway,left,anchor=south,xshift=-3pt,rotate=90] {$\LOGbft{i}{t}$};
        
        \draw [decorate,decoration={brace,amplitude=5pt}]
        ([xshift=-3pt]bs3.south west) -- ([xshift=-3pt]bs1.north west)
        node [midway,left,anchor=south,xshift=-3pt,rotate=90] {$\LOGlc{i}{t}$};
        
        \begin{scope}[yshift=0.0cm]
            \coordinate (bsgen) at (0,0.2);
            \node (bs1) at (0,0) [smallblock] {1};
            
            \draw [largelink] (bs1) -- (bsgen);
        \end{scope}
        
        \begin{scope}[yshift=-1.05cm]
            \coordinate (bsgen) at (0,0.2);
            \node (bs1) at (0,0) [smallblock] {1};
            \node (bs2) at (0,-0.33) [smallblock] {2};
            
            \draw [largelink] (bs2) -- (bs1);
            \draw [largelink] (bs1) -- (bsgen);
        \end{scope}
        
        \begin{scope}[yshift=-2.25cm]
            \coordinate (bsgen) at (0,0.2);
            \node (bs1) at (0,0) [smallblock] {1};
            \node (bs2) at (0,-0.33) [smallblock] {2};
            
            \draw [largelink] (bs2) -- (bs1);
            \draw [largelink] (bs1) -- (bsgen);
        \end{scope}
        
        \draw [-Latex,shorten <=0.5em,shorten >=0.85em] (bl1) -- ++(1,0);
        \draw [-Latex,shorten <=0.5em,shorten >=0.85em] (bl2) -- ++(1,0) node [midway,xshift=-4pt,fill=white,rotate=90] {\scriptsize\emph{Flatten}};
        \draw [-Latex,shorten <=0.5em,shorten >=0.85em] (bl3) -- ++(1,0);
        
        \begin{scope}[yshift=0.0cm,xshift=1.8cm]
            \coordinate (bsgen) at (0,0.2);
            \node (bs1) at (0,0) [smallblock] {1};
        \end{scope}
        
        \begin{scope}[yshift=-1.05cm,xshift=1.8cm]
            \node (bs1) at (0,0) [smallblock] {1};
            \node (bs2) at (0,-0.33) [smallblock] {2};
            
            \draw [red,thick,shorten <=-0.2em,shorten >=-0.2em] (bs1.north west) -- (bs1.south east);
            \draw [red,thick,shorten <=-0.2em,shorten >=-0.2em] (bs1.south west) -- (bs1.north east);
        \end{scope}
        
        \begin{scope}[yshift=-2.25cm,xshift=1.8cm]
            \node (bs1) at (0,0) [smallblock] {1};
            \node (bs2) at (0,-0.33) [smallblock] {2};
            
            \draw [red,thick,shorten <=-0.2em,shorten >=-0.2em] (bs1.north west) -- (bs1.south east);
            \draw [red,thick,shorten <=-0.2em,shorten >=-0.2em] (bs1.south west) -- (bs1.north east);
            
            \draw [red,thick,shorten <=-0.2em,shorten >=-0.2em] (bs2.north west) -- (bs2.south east);
            \draw [red,thick,shorten <=-0.2em,shorten >=-0.2em] (bs2.south west) -- (bs2.north east);
        \end{scope}

        \draw [decorate,decoration={brace,amplitude=5pt,aspect=0.85}]
        ($(bl1.north east) + (0.9,0)$) -- ($(bl3.south east) + (0.9,0)$);
        
        \draw [-Latex,shorten >=0.85em] (1.35,-2) -- (1.75,-2) node [midway,xshift=-4pt,yshift=2.5em,rotate=90] {\scriptsize\emph{Sanitize}}; %

        \begin{scope}[xshift=-0.4cm]
        
            \begin{scope}[yshift=-2.25cm,xshift=3.6cm]
                \node (bs1) at (0,0) [smallblock] {1};
                \node (bs2) at (0,-0.33) [smallblock] {2};
            \end{scope}
            
            \draw [-Latex,shorten >=0.85em] (2.15,-2) -- (2.5,-2);
            
            \begin{scope}[yshift=-2.25cm,xshift=4.5cm]
                \node (bs1) at (0,0) [smallblock,alias=br11] {1};
                \node (bs2) at (0,-0.33) [smallblock] {2};
            \end{scope}
            
            \begin{scope}[xshift=4.5cm]
                \node (bs1) at (0,-3) [smallblock] {1};
                \node (bs2) at (0,-3.33) [smallblock] {2};
                \node (bs3) at (0,-3.66) [smallblock,alias=br12] {3};
                
                \draw [red,thick,shorten <=-0.2em,shorten >=-0.2em] (bs1.north west) -- (bs1.south east);
                \draw [red,thick,shorten <=-0.2em,shorten >=-0.2em] (bs1.south west) -- (bs1.north east);
                
                \draw [red,thick,shorten <=-0.2em,shorten >=-0.2em] (bs2.north west) -- (bs2.south east);
                \draw [red,thick,shorten <=-0.2em,shorten >=-0.2em] (bs2.south west) -- (bs2.north east);
            \end{scope}
            
            \draw [decorate,decoration={brace,amplitude=5pt,aspect=0.78}]
            ($(br11.north east) + (0.1,0)$) -- ($(br12.south east) + (0.1,0)$);
            
            \begin{scope}[xshift=-0.4cm]
            
            \begin{scope}[xshift=6.3cm]
                \node (bs1) at (0,-3) [smallblock] {1};
                \node (bs2) at (0,-3.33) [smallblock] {2};
                \node (bs3) at (0,-3.66) [smallblock,alias=br12] {3};
            \end{scope}
            
            \node (LOGda) at (3.5,-4.5) {$\LOGda{i}{t}$};
            \draw [-Latex,dashed,shorten <=3.9em] (3.5,-2.9) -- (LOGda);
            
            \end{scope}
            
            \draw [-Latex,shorten >=0.85em,shorten <=1em] (0.2,-3.33) -- (2.5,-3.33);
            
            \draw [-Latex,shorten >=0.85em,shorten <=1em] (2.7,-3.33) -- (3.25,-3.33) node [midway,xshift=0pt,yshift=2.5em,rotate=90] {\scriptsize\emph{Sanitize}}; %
            
            \node (LOGfin) at (2,-4.5) {$\LOGfin{i}{t}$};
            \draw [-Latex,dashed,shorten <=2em] (2,-2) -- (LOGfin);
        
        \end{scope}
        
        \end{tikzpicture}%
    }
    \caption[]{%
        Illustration of inner workings of \sac protocols.
        Variants of $\LOG{}{}$ with subscript $i$ and superscript $t$
        denote ledgers as viewed by node $i$ at time $t$.%
    }
    \label{fig:protocol-overview-abstract-and-ledger-extraction} 
\end{figure}

\Sac protocols are constructed
from an off-the-shelf dynamically available protocol $\PIlc$,
\eg, a longest chain (LC) protocol,
and an off-the-shelf partially synchronous BFT protocol $\PIbft$
(see Figure~\ref{fig:protocol-overview-eaf}).
Nodes execute the \sac protocol $\PIsac$
by executing the two sub-protocols in parallel.
The $\PIlc$ sub-protocol receives transactions $\Txs{}{}$ from the environment and outputs an ever-increasing ledger $\LOGlc{}{}$
of transactions.
Over time, each node takes \emph{snapshots} of this ledger based on its own current view, and inputs these snapshots into the second sub-protocol $\PIbft{}{}$.
The output ledger $\LOGbft{}{}$ of $\PIbft$ is an ever-increasing ordered list of such snapshots,
\ie, of \emph{prefixes} of $\LOGlc{}{}$.
To create the finalized ledger $\LOGfin{}{}$ of transactions
(see Figure~\ref{fig:ledger-extraction}),
$\LOGbft{}{}$ is flattened (\ie, all snapshots
included in $\LOGbft{}{}$ are concatenated as ordered)
and sanitized
(\ie, all but the first occurrence of each block are removed).
Finally, $\LOGfin{}{}$ is prepended to $\LOGlc{}{}$ and sanitized to form the available ledger $\LOGda{}{}$.
This ensures the desired prefix property,
namely that all users irrespective of
whether they follow $\LOGda{}{}$ or $\LOGfin{}{}$
eventually agree on a single account of history.

An adversary could, in an attempt to break safety,
input into $\PIbft$
an ostensible snapshot of the ledger $\LOGlc{}{}$
which really contains unconfirmed transactions.
To rule out this possibility,
each honest node
boycotts in $\PIbft$ the finalization
of snapshots that are not locally viewed as confirmed in $\PIlc$.
This requires a minor modification to the BFT protocol,
which is provided in \cite{ebbandflow} for PBFT \cite{pbft}, Hotstuff \cite{yin2018hotstuff} and Streamlet \cite{streamlet}.

When any of these slightly modified BFT protocols
is used in conjunction
with a permissioned LC protocol
\cite{sleepy,david2018ouroboros},
the resulting \sac protocol satisfies the desired goal
of a secure \eaf protocol with optimal resilience:
\begin{theorem*}[Informal version of Theorem 1 in \cite{ebbandflow}]
Consider
a network environment where:
\begin{enumerate}
    \item Communication is asynchronous until a  global stabilization time $\GST$ after which communication becomes synchronous, and 
    \item honest nodes sleep and wake up until a global awake time $\GOT$ after which all nodes are awake. 
    Adversary nodes are always awake.
\end{enumerate}
Then
\begin{enumerate}
    \item \textbf{P1 -- Finality}: The finalized ledger $\LOGfin{}{}$ is guaranteed to be safe at all times, and live after $\max\{\GST,\GOT\}$, provided that fewer than $33\%$ of all the nodes are adversarial.
    \item \textbf{P2 -- Dynamic Availability}: If $\GST = 0$, the available ledger $\LOGda{}{}$ is guaranteed to be safe and live at all times, provided that at all times fewer than $50\%$ of the awake nodes are adversarial.
    \item \textbf{Prefix}: Under all circumstances, $\LOGfin{}{}$ is a prefix of $\LOGda{}{}$.
\end{enumerate}
\end{theorem*}

The proof %
can be found in \cite{ebbandflow}. Here we give some intuition why this theorem is true, giving some insights into the design of \sac protocols:
\begin{itemize}
    \item Under \textbf{P1}, which is the classical partially synchronous environment, the BFT sub-protocol $\PIbft$ is safe. Hence, the finalized ledger $\LOGfin{}{}$, being the sanitized and flattened output of $\PIbft$ is safe as well. This holds regardless of whether the dynamically available protocol $\PIlc$ is safe. Moreover,  both the sub-protocols $\PIlc$ and $\PIbft$ are safe and live after $\max\{\GST.\GOT\}$. Hence the finalized ledger $\LOGfin{}{}$ is safe and live as well after this time. 
    \item Under \textbf{P2}, the dynamically available sub-protocol $\PIlc$ is safe and live. Regardless of whether $\PIbft$ is safe, the snapshots that are fed into  $\PIbft$ are prefixes of each other and whatever $\PIbft$ finalizes is always consistent with the output ledger $\LOGlc{}{}$ of $\PIlc$. Thus, the finalized ledger $\LOGfin{}{}$ is a prefix of $\LOGlc{}{}$, and the available ledger $\LOGda{}{}{}{}$, obtained by prepending $\LOGfin{}{}$ to $\LOGlc{}{}$, is just $\LOGlc{}{}$, and is hence safe and live. 
\end{itemize}

\paragraph{Comparison to Finality Gadget Based Designs}

An alternative approach to obtain \eaf protocols is to design a finality gadget (FG) \cite{buterin2017casper,fg_nightshade,dinsdale2019afgjort,ethresearch-hierarchical-finality-gadget,stewart2020grandpa} in conjunction with a dynamically available LC protocol as block proposal mechanism.
The key difference between \sac protocols and typical FG-based designs
is how consistency between the available and the finalized ledger is maintained.
For FG-based designs,
the ledger $\LOGlc{}{}$ of the dynamically available block proposal mechanism %
and the ledger $\LOGbft{}{}{}$ of the FG %
are output unaltered as $\LOGda{}{}$ and $\LOGfin{}{}$, respectively.
To ensure that $\LOGfin{}{}$ is always a prefix of $\LOGda{}{}$,
the fork choice rule of the dynamically available protocol %
has to be
modified to honor prior finalizations.
This feedback from the FG into the LC protocol
renders security arguments intricate and
can be a gateway for attacks \cite{ethresearch-bouncing-attack,ethresearch-bouncing-attack-analysis}.
\Sac protocols, on the other hand,
compute $\LOGda{}{}$ and $\LOGfin{}{}$
from $\LOGlc{}{}$ and $\LOGbft{}{}{}$
via a ledger extraction procedure
(Figure~\ref{fig:ledger-extraction})
that absorbs any inconsistencies that may arise.
The modular design and absence of feedback from $\PIbft$ to $\PIlc$
facilitates a stringent security proof.

\hyphenation{syn-chro-nous}

What is more,
the modular construction of \sac protocols
allows the use of state-of-the-art dynamically available protocols and state-of-the-art partially synchronous BFT protocols
without the need to reinvent the wheel.
As a result, \sac protocols
can take advantage of future advances in the design of dynamically available protocols and in the design of partially synchronous BFT protocols.
Furthermore, \sac protocols can readily take advantage of additional features
provided by its unmodified sub-protocols.
This strength is showcased in Section~\ref{sec:accountability}
where a cryptoeconomic security property (accountable safety)
of HotStuff and Streamlet is readily inherited by \sac
protocols using either of the two BFT protocols as $\PIbft$.
In contrast, FG-based designs entail handcrafting
of the finality voting, the fork choice rule of 
the underlying Nakamoto-style chain,
and all additional required features.

However,
as a result of how they maintain consistency between the ledgers,
FG-based designs and \sac protocols differ
when it comes to determining transaction validity.
In an FG-based design, both ledgers come
from a single chain in the underlying LC protocol.
Hence, if a transaction $\tx$ becomes part of one of the ledgers
via an \blocklc $b$, then the transactions that precede $\tx$
in the ledger
are those in the prefix of $b$.
Hence, the validity of $\tx$ can be determined
at the time $b$ is produced.
As a consequence, it is a typical requirement
that valid blocks only contain valid transactions.
This avoids wasting resources on invalid transactions
and is leveraged in standard light client constructions.

On the other hand,
in \sac protocols %
it is possible for two snapshots
in $\PIbft$ to come from two conflicting LC chains
that were produced, \eg, during a period of asynchrony.
In this case, transactions from the first LC chain
enter into $\LOGfin{}{}$ %
before transactions from the second LC chain,
potentially invalidating some of the latter
although they seemed valid at the time
the containing \blocklc $b$ was composed
considering the prefix of $b$.
Consequently,
unlike for FG-based constructions,
it is not possible to determine with certainty
at the time of composing an \blocklc $b$
whether a contained transaction $b$
will be considered valid
when
it is finally inserted into $\LOGfin{}{}$.\footnote{%
Reorganization of transactions as they
enter $\LOGfin{}{}$ occurs only in the context of
network partitions. Once honest validators have caught up
with each other after a partition, no further
reorganizations take place and nodes can predict
the validity of transactions
-- until the next network partition.}
Since $\LOGfin{}{}$ is prepended to $\LOGlc{}{}$ to obtain $\LOGda{}{}$,
the same argument applies to $\LOGda{}{}$.
While the non-trivial ledger extraction of \sac protocols
allows to decouple the sub-protocols and thus enables
a stringent security proof for \sac protocols,
it poses challenges when it comes to light client
support, as standard techniques cannot be applied readily.
Nonetheless,
in Section~\ref{sec:light-clients}
we present a twist %
to enable light clients for \sac protocols.

%% file: accountability.tex
\section{Accountability}
\label{sec:accountability}

A typical partially synchronous BFT protocol
is secure
as long as less than a third of validators are adversarial,
and this threshold
is optimal \cite{model-psync}.
But what if more than a third of validators deviate from
the protocol?
A safety violation can then not be avoided.
However,
\emph{accountable safety} provides a
\emph{cryptoeconomic notion of security},
where a safety violation cannot be prevented,
but at least it can be attributed irrefutably
to the wrongdoing of the validators that have caused it,
and in response these validators' stake can be confiscated.
With such a retribution mechanism in place,
a safety violation comes at a cost to the malicious actors,
leading to better incentive alignment of protocol participants.

Following \cite{buterin2017casper,buterin2020combining},
we develop accountability around a set of
\emph{slashing conditions} that
a malicious validator must violate to produce a safety violation
(in which case the misconduct will be observable to honest validators
that suffer from the safety violation)
and that an honest validator will never violate.
\begin{definition}
A protocol provides \emph{$\alpha$-accountable safety} iff:
\begin{enumerate}
    \item In the case of a safety violation of the produced ledger,
        (at least) $\alpha \cdot n$ validators can be irrefutably accused of protocol violations.
        (In particular, they must have violated
        one of the protocol's slashing conditions.)
    \item An honest validator cannot be credibly falsely accused.
        (In particular, honest validators never violate the protocol's slashing conditions.)
\end{enumerate}
\end{definition}

We show that both Streamlet \cite{streamlet}
and HotStuff \cite{yin2018hotstuff} provide
$\frac{1}{3}$-accountable safety.
Since a necessary condition for a safety violation of $\LOGfin{}{}$
of a \sac protocol is a safety violation of $\PIbft$,
the following readily follows:

\begin{theorem}
\Sac protocols, when constructed using Streamlet or HotStuff as $\PIbft$,
provide $\frac{1}{3}$-accountable safety for $\LOGfin{}{}$.
\end{theorem}

This showcases a strength of the almost black-box construction
of \sac protocols.
Some properties and features of the constituent protocols can readily
be inherited by the resulting \sac protocol.

First, we identify candidate slashing conditions for Streamlet:
\begin{definition}[Streamlet slashing conditions]
Let $B$ denote a BFT block at depth $|B|$, produced in Streamlet epoch $e_B$.
A validator's stake is slashed if:
\begin{enumerate}
    \item
        The validator
        votes for $B_1, B_2$ such that
        $e_{B_1} = e_{B_2}$.
    \item
        The validator
        votes for $B_1, B_2$ such that
        $e_{B_1} < e_{B_2}$ but $|B_1| > |B_2|$.
\end{enumerate}
\end{definition}

Accountable safety (with parameter $\alpha$) has two complementary aspects, which we prove subsequently (for $\alpha=\frac{1}{3}$):
\begin{lemma}
\label{thm:accountability-streamlet-attack}
In the case of a safety violation in Streamlet,
we can pinpoint $\alpha \cdot n$ validators that must have violated
a slashing condition.
\end{lemma}
\begin{lemma}
\label{thm:accountability-streamlet-defense}
Honest validators never violate a slashing condition in Streamlet.
\end{lemma}

\tikzset{blockchain/.style={
        x=0.3cm,
        y=0.5cm,
        node distance=0.5cm,
        block/.style = {
            minimum width=0.25cm,
            minimum height=0.25cm,
            draw,
            shade,
            top color=white,
            bottom color=black!10,
        },
        link/.style = {
        },
    }
}

\begin{figure}[tb]
    \centering
    \begin{tikzpicture}
        \small
        
        \begin{scope}[blockchain]
        
            \coordinate (b0) at (0,-0.25);
            \coordinate (b11) at (0,-1);
            \coordinate (b21) at (-1,-2);
            \node (b22) at (+1,-2) [block] {};
            \node (b31) at (-1,-3) [block] {};
            \node (b32) at (+1,-3) [block] {};
            \node (b42) at (+1,-4) [block] {};
            \node (b51) at (-1,-5) [block] {};
            \node (b61) at (-1,-6) [block] {};
            \node (b71) at (-1,-7) [block] {};
            
            \draw [link,densely dotted] (b11) -- (b0);
            \draw [link,densely dotted] (b21) -- (b11);
            \draw [link,densely dotted] (b22) -- (b11);
            \draw [link] (b32) -- (b22);
            \draw [link] (b42) -- (b32);
            \draw [link] (b61) -- (b51);
            \draw [link] (b71) -- (b61);
            \draw [link,densely dotted] (b51) -- (b31);
            \draw [link,densely dotted] (b31) -- (b21);
            
            \node [anchor=east,left=0.3cm of b31] {$B$} node [anchor=east,left=1cm of b31] {$e_B$};
            
            \node [anchor=east,left=0.3cm of b51] {$B_1'$} node [anchor=east,left=1cm of b51] {$(e'-1)$};
            \node [anchor=east,left=0.3cm of b61] {$B_2'$} node [anchor=east,left=1cm of b61] {$e'$};
            \node [anchor=east,left=0.3cm of b71] {$B_3'$} node [anchor=east,left=1cm of b71] {$(e'+1)$};
            
            \node [anchor=west,right=0.3cm of b22] {$B_1$} node [anchor=west,right=1cm of b22] {$(e-1)$};
            \node [anchor=west,right=0.3cm of b32] {$B_2$} node [anchor=west,right=1cm of b32] {$e$};
            \node [anchor=west,right=0.3cm of b42] {$B_3$} node [anchor=west,right=1cm of b42] {$(e+1)$};
            
        \end{scope}

    \end{tikzpicture}%
    \caption{Safety violation in Streamlet}
    \label{fig:accountability-streamlet-safety-violation}
\end{figure}
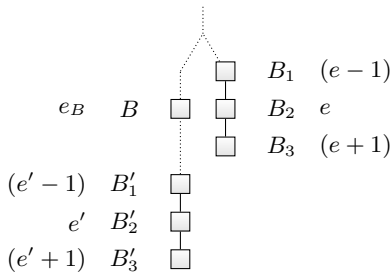

The proof of Lemma~\ref{thm:accountability-streamlet-attack}
proceeds along the safety argument of Streamlet.
\begin{proof}
Suppose there is a safety violation (see Figure~\ref{fig:accountability-streamlet-safety-violation}),
\ie,
from blocks $B_1, B_2, B_3$ of epochs $e-1, e, e+1$
and blocks $B'_1, B'_2, B'_3$ of epochs $e'-1, e', e'+1$,
finalizing conflicting blocks $B_2, B'_2$.
Without loss of generality, let $|B_2| \leq |B'_2|$.
Then there must be a block $B \neq B_2$ with $|B| = |B_2|$,
let it be from epoch $e_B$. All blocks must be notarized,
\ie, $\geq 2n/3$ validators have voted for each block.

If $e_B \in \{e-1, e, e+1\}$, then by quorum intersection
$\geq 1/3$ of validators have voted for both $B$
and one of $\{B_1, B_2, B_3\}$ in the same epoch $e_B$.
A violation of the first slashing condition.

If $e_B < e-1$, then by quorum intersection
$\geq 1/3$ of validators have first voted in epoch $e_B$ for $B$ of height $|B|$
and at a later epoch $e-1$ voted for $B_1$ of lesser height $|B_1| = |B| - 1 < |B|$.
A violation of the second slashing condition.

If $e+1 < e_B$, then by quorum intersection
$\geq 1/3$ of validators have first voted in epoch $e+1$ for $B_3$ of height $|B_3|$
and at a later epoch $e_B$ voted for $B$ of lesser height $|B| = |B_3| - 1 < |B_3|$.
A violation of the second slashing condition.
\end{proof}

The proof of Lemma~\ref{thm:accountability-streamlet-defense}
follows by inspecting the Streamlet pseudocode \cite{streamlet}.
\begin{proof}
        An honest validator does not violate the first
        slashing condition
        because it only votes once per epoch.
        An honest validator does not violate the second
        slashing condition
        because if it has had
        a notarized chain in view
        of length $|B_1|-1$ in epoch $e_{B_1}$
        (otherwise it would not have voted for $B_1$),
        then it will not vote for $B_2$
        of depth $|B_2| \leq |B_1|-1 < |B_1|$ in a later
        epoch $e_{B_2}$ because $B_2$ does not extend the
        longest notarized chain
        (which is at least of length $|B_1|-1$).
\end{proof}

From Lemmas~\ref{thm:accountability-streamlet-attack} and \ref{thm:accountability-streamlet-defense}
readily follows:
\begin{theorem}
Partially synchronous Streamlet provides $\frac{1}{3}$-accountable safety.
\end{theorem}

Thanks to the similarities between Streamlet and HotStuff,
a very similar argument establishes accountable safety for HotStuff.
Details in Appendix~\ref{sec:appendix-accountability-hotstuff}.
An independent contemporaneous work \cite{kannan2020bft}
reaches a similar conclusion and
in particular details how to acquire proof of
slashing condition violations.

%% file: light_clients.tex
\section{Light Clients and Simple Payment Verification}
\label{sec:light-clients}

\tikzset{blockchain/.style={
        x=0.3cm,
        y=0.6cm,
        node distance=0.5cm,
        block/.style = {
            minimum width=0.25cm,
            minimum height=0.25cm,
            draw,
            shade,
            top color=white,
            bottom color=black!10,
        },
        link/.style = {
        },
        refsnapshot/.style = {
            {Circle[length=3pt]}-latex,
            myParula01Blue,
            bend right=30,
            shorten <=-1.5pt,
            thick,
        },
        refLOGfin/.style = {
            {Circle[length=3pt]}-latex,
            myParula02Orange,
            bend left=10,
            shorten <=-1.5pt,
            thick,
        },
        refsnapshotLIGHT/.style = {
            refsnapshot,
            myParula01Blue!20,
            thin,
        },
        refLOGfinLIGHT/.style = {
            refLOGfin,
            myParula02Orange!20,
            thin,
        },
    }
}

Simple payment verification by light clients \cite{nakamoto_paper}
(which only follow the blockchain headers but not the transaction data)
can be supported
after adding a few metadata
to LC and BFT blocks
and extending the block validity rules to ensure the
proper calculation of these auxiliary fields.
This metadata serves as a trust anchor with respect to which
full nodes can answer light clients' queries verifiably.

The goal is to design simple payment verification such that
if a light client obtains an answer to its query, then
a full client would have received the same answer.
In some circumstances where the full client can obtain an answer
(but only one that is potentially invalidated in the future
due to a safety violation in $\LOGda{}{}$),
the light client's query might not be answerable in the given
state from the given metadata, and the light client is asked
to check back later.

In the following, we first describe the additional metadata
added to BFT and LC blocks to enable light clients.
Then, we describe the process of simple payment verification (SPV),
which is the light client algorithm.
Light clients use data availability proofs
\cite{dankrad-data-availability,albassam2018fraud,codedmerkletree}
before accepting a block's header
to ensure that the block is available for full nodes
to answer light client queries.

\paragraph{BFT Blocks}

When a validator proposes a BFT block $B$
extending $B'$, it knows the prefix
$\LOGfin{}{}(B')$ the block $B$ will have
once it appears in $\LOGfin{}{}$
(and also in $\LOGda{}{}$).
It can thus extract the sequence of transactions
$\Delta\Txs{}{}(B)$ that are newly introduced by $B$
into the ledgers after sanitization,
\ie, $\Delta\Txs{}{}(B)$ is the sequence of transactions
such that
\begin{IEEEeqnarray}{C}
    \LOGfin{}{}(B) = \LOGfin{}{}(B') \| \Delta\Txs{}{}(B).
\end{IEEEeqnarray}
We refer to $\Delta\Txs{}{}(B)$ as the
\emph{innovation} brought by $B$.
The BFT block producer commits to $\Delta\Txs{}{}(B)$
using, \eg, a Merkle tree \cite{merkle},
the root of which is put into the header of $B$.
All other validators can verify that the Merkle root was computed
correctly and only then consider $B$ valid.

\paragraph{LC Blocks}

\begin{figure}[tb!]
    \centering
    \subfloat[The latest BFT block is the $\protect\LOGfin{}{}$ reference included in the most recent LC block -- SPV can proceed.\label{fig:lightclient-1-trivial}]{%
        \begin{minipage}[t]{0.45\linewidth}\centering%
        \begin{tikzpicture}
            \small
            
            \begin{scope}[blockchain,xshift=-1cm]
            
                \node (label) at (-4,-1) {LC};
            
                \coordinate (b0) at (0,-0.5);
                \node (b11) at (0,-1) [block] {};
                \node (b21) at (-1,-2) [block] {};
                \node (b22) at (+1,-2) [block] {};
                \node (b32) at (+1,-3) [block] {};
                \node (b42) at (+1,-4) [block] {};
                \node (b52) at (+1,-5) [block] {};
                \node (b62) at (+1,-6) [block] {};
                
                \draw [link] (b11) -- (b0);
                \draw [link] (b21) -- (b11);
                \draw [link] (b22) -- (b11);
                \draw [link] (b32) -- (b22);
                \draw [link] (b42) -- (b32);
                \draw [link] (b52) -- (b42);
                \draw [link] (b62) -- (b52);
                
            \end{scope}
            
            \begin{scope}[blockchain,xshift=+1cm]
            
                \node (label) at (3,-1) {BFT};
                
                \coordinate (B0) at (0,-0.5);
                \node (B11) at (0,-1) [block] {};
                \node (B21) at (0,-2) [block] {};
                \node (B31) at (0,-3) [block] {};
                \node (B41) at (0,-4) [block] {};
                
                \draw [link] (B11) -- (B0);
                \draw [link] (B21) -- (B11);
                \draw [link] (B31) -- (B21);
                \draw [link] (B41) -- (B31);
                
            \end{scope}
            
            \begin{scope}[blockchain]
            
                \draw [refsnapshotLIGHT] (B11.center) to (b11);
                \draw [refsnapshotLIGHT] (B21.center) to (b21);
                \draw [refsnapshotLIGHT] (B31.center) to (b32);
                
                \draw [refLOGfinLIGHT] (b52.center) to (B21);
                \draw [refLOGfinLIGHT] (b42.center) to (B21);
                \draw [refLOGfinLIGHT] (b42.center) to (B21);
                \draw [refLOGfinLIGHT] (b32.center) to (B11);
                
                \draw [refsnapshot] (B41.center) to (b42);
                \draw [refLOGfin] (b62.center) to (B41);
                
            \end{scope}
            
        \end{tikzpicture}%
        \end{minipage}
    }\hfill%
    \subfloat[Snapshots introduced by BFT blocks since the last $\protect\LOGfin{}{}$ reference are consistent with the most recent LC block -- SPV can proceed.\label{fig:lightclient-2-benign}]{%
        \begin{minipage}[t]{0.45\linewidth}\centering%
        \begin{tikzpicture}
            \small
            
            \begin{scope}[blockchain,xshift=-1cm]
            
                \node (label) at (-4,-1) {LC};
            
                \coordinate (b0) at (0,-0.5);
                \node (b11) at (0,-1) [block] {};
                \node (b21) at (-1,-2) [block] {};
                \node (b22) at (+1,-2) [block] {};
                \node (b32) at (+1,-3) [block] {};
                \node (b42) at (+1,-4) [block] {};
                \node (b52) at (+1,-5) [block] {};
                \node (b62) at (+1,-6) [block] {};
                
                \draw [link] (b11) -- (b0);
                \draw [link] (b21) -- (b11);
                \draw [link] (b22) -- (b11);
                \draw [link] (b32) -- (b22);
                \draw [link] (b42) -- (b32);
                \draw [link] (b52) -- (b42);
                \draw [link] (b62) -- (b52);
                
            \end{scope}
            
            \begin{scope}[blockchain,xshift=+1cm]
            
                \node (label) at (3,-1) {BFT};
                
                \coordinate (B0) at (0,-0.5);
                \node (B11) at (0,-1) [block] {};
                \node (B21) at (0,-2) [block] {};
                \node (B31) at (0,-3) [block] {};
                \node (B41) at (0,-4) [block] {};
                \node (B51) at (0,-5) [block] {};
                \node (B61) at (0,-6) [block] {};
                
                \draw [link] (B11) -- (B0);
                \draw [link] (B21) -- (B11);
                \draw [link] (B31) -- (B21);
                \draw [link] (B41) -- (B31);
                \draw [link] (B51) -- (B41);
                \draw [link] (B61) -- (B51);
                
            \end{scope}
            
            \begin{scope}[blockchain]
            
                \draw [refsnapshotLIGHT] (B11.center) to (b11);
                \draw [refsnapshotLIGHT] (B21.center) to (b21);
                \draw [refsnapshotLIGHT] (B31.center) to (b32);
                
                \draw [refLOGfinLIGHT] (b52.center) to (B21);
                \draw [refLOGfinLIGHT] (b42.center) to (B21);
                \draw [refLOGfinLIGHT] (b42.center) to (B21);
                \draw [refLOGfinLIGHT] (b32.center) to (B11);
                
                \draw [refsnapshot] (B41.center) to (b42);
                \draw [refLOGfin] (b62.center) to (B41);
                
                \draw [refsnapshot,bend left=10] (B51.center) to (b32);
                \draw [refsnapshot] (B61.center) to (b52);
                
                \node [right of=B51] {\textcolor{myParula05Green}{\cmark}};
                \node [right of=B61] {\textcolor{myParula05Green}{\cmark}};
                
            \end{scope}
            
        \end{tikzpicture}%
        \end{minipage}
    }\\%
    \subfloat[Snapshots introduced by BFT blocks since the last $\protect\LOGfin{}{}$ reference are \emph{not} consistent with the most recent LC block -- SPV is unavailable.\label{fig:lightclient-3-easyproblem}]{%
        \begin{minipage}[t]{0.45\linewidth}\centering%
        \begin{tikzpicture}
            \small
            
            \begin{scope}[blockchain,xshift=-1cm]
            
                \node (label) at (-4,-1) {LC};
            
                \coordinate (b0) at (0,-0.5);
                \node (b11) at (0,-1) [block] {};
                \node (b21) at (-1,-2) [block] {};
                \node (b22) at (+1,-2) [block] {};
                \node (b31) at (-1,-3) [block] {};
                \node (b32) at (+1,-3) [block] {};
                \node (b41) at (-1,-4) [block] {};
                \node (b42) at (+1,-4) [block] {};
                \node (b52) at (+1,-5) [block] {};
                \node (b62) at (+1,-6) [block] {};
                
                \draw [link] (b11) -- (b0);
                \draw [link] (b21) -- (b11);
                \draw [link] (b22) -- (b11);
                \draw [link] (b31) -- (b21);
                \draw [link] (b32) -- (b22);
                \draw [link] (b41) -- (b31);
                \draw [link] (b42) -- (b32);
                \draw [link] (b52) -- (b42);
                \draw [link] (b62) -- (b52);
                
            \end{scope}
            
            \begin{scope}[blockchain,xshift=+1cm]
            
                \node (label) at (3,-1) {BFT};
                
                \coordinate (B0) at (0,-0.5);
                \node (B11) at (0,-1) [block] {};
                \node (B21) at (0,-2) [block] {};
                \node (B31) at (0,-3) [block] {};
                \node (B41) at (0,-4) [block] {};
                \node (B51) at (0,-5) [block] {};
                \node (B61) at (0,-6) [block] {};
                
                \draw [link] (B11) -- (B0);
                \draw [link] (B21) -- (B11);
                \draw [link] (B31) -- (B21);
                \draw [link] (B41) -- (B31);
                \draw [link] (B51) -- (B41);
                \draw [link] (B61) -- (B51);
                
            \end{scope}
            
            \begin{scope}[blockchain]
            
                \draw [refsnapshotLIGHT] (B11.center) to (b11);
                \draw [refsnapshotLIGHT] (B21.center) to (b21);
                \draw [refsnapshotLIGHT] (B31.center) to (b32);
                
                \draw [refLOGfinLIGHT] (b52.center) to (B21);
                \draw [refLOGfinLIGHT] (b42.center) to (B21);
                \draw [refLOGfinLIGHT] (b42.center) to (B21);
                \draw [refLOGfinLIGHT] (b32.center) to (B11);
                
                \draw [refsnapshot] (B41.center) to (b42);
                \draw [refLOGfin] (b62.center) to (B41);
                
                \draw [refsnapshot,bend left=15] (B51.center) to (b41);
                \draw [refsnapshot,bend right=10] (B61.center) to (b52);
                
                \node [right of=B51] {\textcolor{myParula07Red}{\xmark}};
                \node [right of=B61] {\textcolor{myParula05Green}{\cmark}};
                
            \end{scope}
            
        \end{tikzpicture}%
        \end{minipage}
    }\hfill%
    \subfloat[Complicated example of fictional ledger reorganization during which it is unsafe to use SPV because latest snapshots could introduce inconsistencies.\label{fig:lightclient-4-harderproblem}]{%
        \begin{minipage}[t]{0.45\linewidth}\centering%
        \begin{tikzpicture}
            \small
            
            \begin{scope}[blockchain,xshift=-1cm]
            
                \node (label) at (-3,-1) {LC};
            
                \coordinate (b0) at (0,-0.5);
                \node (b11) at (0,-1) [block] {};
                \node (b21) at (-1,-2) [block] {};
                \node (b22) at (+1,-2) [block] {};
                \node (b31) at (-1,-3) [block] {};
                \node (b32) at (+1,-3) [block] {};
                \node (b41) at (-1,-4) [block] {};
                \node (b42) at (+1,-4) [block] {};
                \node (b51) at (-1,-5) [block] {};
                \node (b52) at (+1,-5) [block] {};
                \node (b62) at (+1,-6) [block] {};
                
                \draw [link] (b11) -- (b0);
                \draw [link] (b21) -- (b11);
                \draw [link] (b22) -- (b11);
                \draw [link] (b31) -- (b21);
                \draw [link] (b32) -- (b22);
                \draw [link] (b41) -- (b31);
                \draw [link] (b42) -- (b32);
                \draw [link] (b51) -- (b41);
                \draw [link] (b52) -- (b42);
                \draw [link] (b62) -- (b52);
                
            \end{scope}
            
            \begin{scope}[blockchain,xshift=+2cm,x=0.5cm]
            
                \node (label) at (1.8,-1) {BFT};
                
                \coordinate (B0) at (0,-0.5);
                \node (B11) at (0,-1) [block] {};
                \node (B21) at (0,-2) [block] {};
                \node (B31) at (0,-3) [block] {};
                \node (B41) at (-1,-4) [block] {};
                \node (B42) at (+1,-4) [block] {};
                \node (B51) at (-1,-5) [block] {};
                \node (B52) at (+1,-5) [block] {};
                \node (B62) at (+1,-6) [block] {};
                
                \draw [link] (B11) -- (B0);
                \draw [link] (B21) -- (B11);
                \draw [link] (B31) -- (B21);
                \draw [link] (B41) -- (B31);
                \draw [link] (B42) -- (B31);
                \draw [link] (B51) -- (B41);
                \draw [link] (B52) -- (B42);
                \draw [link] (B62) -- (B52);
                
            \end{scope}
            
            \begin{scope}[blockchain,node distance=0.35cm]
            
                \draw [refsnapshotLIGHT] (B11.center) to (b11);
                \draw [refsnapshotLIGHT] (B21.center) to (b21);
                \draw [refsnapshotLIGHT] (B31.center) to (b32);
                
                \draw [refLOGfinLIGHT] (b52.center) to (B21);
                \draw [refLOGfinLIGHT] (b42.center) to (B21);
                \draw [refLOGfinLIGHT] (b42.center) to (B21);
                \draw [refLOGfinLIGHT] (b32.center) to (B11);
                
                \draw [refLOGfin] (b62.center) to (B51);
                
                \draw [refsnapshot] (B41.center) to (b41);
                \draw [refsnapshot] (B51.center) to (b52);
                
                \draw [refsnapshot] (B42.center) to (b42);
                \draw [refsnapshot,bend right=10] (B52.center) to (b42);
                \draw [refsnapshot,bend left=20] (B62.center) to (b51);

                \node [right of=B41] {\textcolor{myParula07Red}{\xmark}};
                \node [right of=B51] {\textcolor{myParula05Green}{\cmark}};
                \node [right of=B42] {\textcolor{myParula05Green}{\cmark}};
                \node [right of=B52] {\textcolor{myParula05Green}{\cmark}};
                \node [right of=B62] {\textcolor{myParula07Red}{\xmark}};
                
            \end{scope}
            
        \end{tikzpicture}%
        \end{minipage}
    }%
    \caption[]{%
        Whether simple payment verification
        can be conducted by the light client
        depends on whether the latest BFT blocks
        since the $\LOGfin{}{}$ reference
        (\tikz{ \draw [blockchain,refLOGfin,bend left=20] (0,0) to (2,0); })
        included in the latest LC block
        introduce incompatible snapshots
        or not
        (\tikz{ \draw [blockchain,refsnapshot,bend right=20] (0,0) to (-2,0); }).}
    \label{fig:lightclient}
\end{figure}
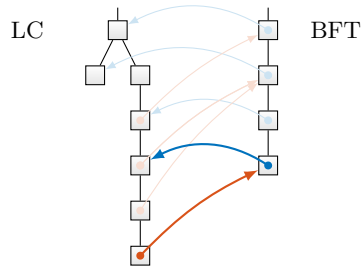
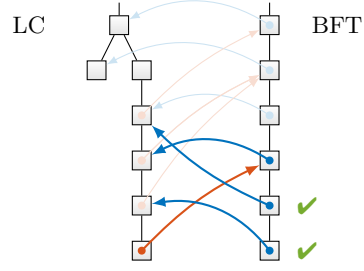
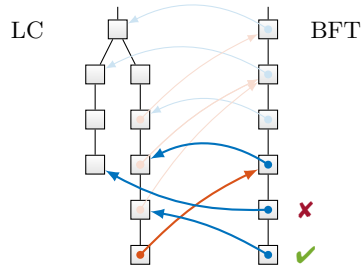
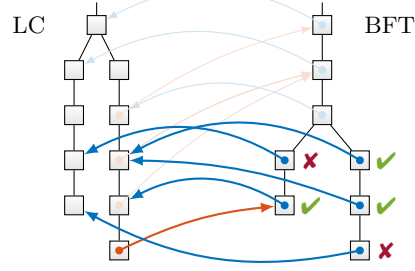

When a validator proposes an LC block $b$
extending $b'$, it knows which prefix
$\LOGlc{}{}(b')$ the block $b$ will have
once it appears in $\LOGlc{}{}$.
Furthermore, the validator $i$ knows what prefix it would
have in $\LOGda{}{}$ if $\LOGlc{}{}$ was prefixed with
the current $\LOGfin{i}{t}$ (\ie, $\LOGfin{}{}$
\emph{as it currently stands}).\footnote{%
To avoid overhead caused by invalid transactions,
the validator should not include
transactions in $b$ that are invalidated by the current prefix,
although the validator cannot be expected to avoid all
conflicts with future prefixes and thus might include
transactions that will be invalidated by later $\LOGfin{}{}$.}
Thus, the validator can extract the \emph{innovation},
\ie, the sequence of transactions
that are introduced by $b$
and its prefix
into $\LOGda{}{}$ after sanitization,
\emph{assuming the current $\LOGfin{}{}$}.
It commits to this sequence, \eg, using a Merkle tree,
the root of which is put into the header of $b$.
Furthermore, the block producer adds to the LC block
a reference
(see Figure~\ref{fig:lightclient},
\tikz{ \draw [blockchain,refLOGfin,bend left=20] (0,0) to (2,0); })
pointing to the BFT block
$B$ representing $\LOGfin{i}{t}$ to provide additional information
on what state of $\LOGfin{}{}$ was underlying
the computation of the Merkle tree.
Other validators can verify that the Merkle root was computed
correctly and only then consider $b$ valid.
Furthermore, they consider $b$ valid only if the referenced
BFT block (representing $\LOGfin{i}{t}$)
is of no smaller depth than the BFT block referenced in $b'$,
the parent of $b$.
This ensures that the reference $\LOGfin{}{}$ gets pushed
forward frequently by the honest validators, and the adversary
cannot downgrade to earlier, grossly outdated,
reference $\LOGfin{}{}$.

\paragraph{Simple Payment Verification (SPV)}

Note that just like full clients,
light clients either believe in
network partitions or not.
Based on this belief they decide
whether to follow $\LOGfin{}{}$ or $\LOGda{}{}$.

If a light client follows $\LOGfin{}{}$,
then the metadata introduced above into the BFT block headers
suffices to succinctly prove to a client
that a certain transaction has been executed at a certain point
in $\LOGfin{}{}$, by pointing the light client to the respective
block $B$
(the header of which the light client knows, including
the Merkle root)
and providing the Merkle proof corresponding to the transaction.
This mechanism is directly analogous to SPV for Bitcoin \cite{nakamoto_paper}.

If a light client follows $\LOGda{}{}$,
then the metadata introduced above into the BFT and LC block headers
can be used to succinctly prove to a client
that a certain transaction has been executed at a certain point
in $\LOGda{}{}$ in most ordinary circumstances.
If the latest confirmed LC block computed its light client
auxiliary information based on a $\LOGfin{}{}$ reference
(\tikz{ \draw [blockchain,refLOGfin,bend left=20] (0,0) to (2,0); })
that corresponds to the latest state of $\LOGfin{}{}$ %
as viewed by the light client
(see Figure~\ref{fig:lightclient-1-trivial}),
then the light client can be convinced that a certain
transaction has taken place by
providing the Merkle proof corresponding to the transaction
relative to a Merkle root in either the LC block
or in one of the BFT blocks in the prefix of the $\LOGfin{}{}$ reference
(depending on where the transaction was introduced).

If new BFT blocks have been finalized, so that $\LOGfin{}{}$ %
as viewed by the light client
is ahead of the $\LOGfin{}{}$ reference of the latest
confirmed LC block (Figure~\ref{fig:lightclient-2-benign}),
but all the additional BFT blocks' snapshots
(\tikz{ \draw [blockchain,refsnapshot,bend right=20] (0,0) to (-2,0); })
lie in the prefix
of the LC block, then the additional BFT blocks are
harmless (\textcolor{myParula05Green}{\cmark}) in the sense that they merely move transactions
from $\LOGda{}{}$ to $\LOGfin{}{}$ but without changing the order
or introducing or removing transactions, and thus
it is safe to rely on the
light client metadata for SPV.

If new BFT blocks have been finalized whose
snapshots lie outside of the prefix of the LC block
(Figure~\ref{fig:lightclient-3-easyproblem}),
then this is dangerous (\textcolor{myParula07Red}{\xmark})
as new transactions were potentially introduced in $\LOGfin{}{}$
that can invalidate some of the transactions in the latest
LC block.
The light client metadata is potentially stale
and cannot be used. %
In this case, the light client is not given a definitive
answer and is instead asked to wait for the next LC block
that will hopefully update the $\LOGfin{}{}$ reference.
Note, that this situation appears only under
network condition/adversary \textbf{P1}
when $\LOGda{}{}$ is undergoing a safety violation to maintain
prefix consistency with $\LOGfin{}{}$. In this case,
$\LOGda{}{}$ is unreliable. Still, we uphold that if
the light client receives an answer to its query, then it must
be the same answer that a full client would have received in the
same situation. As that cannot be provided in this case,
no answer is given to the light client.

The general rule is that the light client is not given
a response to its query if on the path from
the $\LOGfin{}{}$ reference to the current $\LOGfin{}{}$ %
as viewed by the light client
there are snapshots that are not entirely contained in the
prefix of the LC block.
A more complicated problematic case is shown in
Figure~\ref{fig:lightclient-4-harderproblem}.
In the example at hand, moving from the $\LOGfin{}{}$ reference
to the latest finalized BFT block
potentially changes the order in which
transactions from the left/right LC chain
enter into $\LOGfin{}{}$.
Thus, the light client cannot answer
the query using the auxiliary information.
Note that this scenario is fictional as it
involves a safety violation of BFT, which is possible only under
network condition/adversary
\textbf{P2}, under which in turn LC is safe and thus all snapshots
are really fully contained in the prefix of the LC block.
Thus, within the environments considered for \eaf protocols,
this case cannot occur %
and is purely illustrative.

\begin{algorithm}[tb]
    \caption{\footnotesize Pseudocode for light client support of \sac protocols}
    \label{algo:pseudocode-light-client}
    \begin{algorithmic}[1]
        \footnotesize
        \Function{ComposeBftBlock}{$B^*, b^*$}
            \Comment{Input: tips of BFT and LC chain}
            \State $\Delta\Txs{}{} \gets \supersanitize(\LOGfin{}{}(B^*) \| \LOGlc{}{}(b^*)) \concatminus \LOGfin{}{}(B^*)$
            \State \Return $(
                \mathrm{prev}=B^*,
                b=b^*,
                \mathrm{auxinnov}=\merkleize(\Delta\Txs{}{})
            )$
        \EndFunction
        \Function{ComposeLcBlock}{$B^*, b^*, \Txs{\mathrm{new}}{}$}
            \State $\Delta\Txs{}{} \gets \supersanitize(\LOGfin{}{}(B^*) \| \LOGlc{}{}(b^*) \| \Txs{\mathrm{new}}{}) \concatminus \LOGfin{}{}(B^*)$
            \State \Return $(
                \mathrm{prev}=b^*,
                \Txs{}{}=\Txs{\mathrm{new}}{},
                \mathrm{auxref}=B^*,
                \mathrm{auxinnov}=\merkleize(\Delta\Txs{}{})
            )$
        \EndFunction
        \Function{SimplePaymentVerificationForAvailableLedger}{$B^*, b^*$}
            \If{$|\{B \in (\text{path from $b^*.\mathrm{auxref}$ to $B^*$}) \mid B.b \not\preceq b^* \land b^* \not\preceq B.b\}|>0$}
                \State \Return $\mathrm{\AcceptMerkleProofWrt}(\{ B.\mathrm{auxinnov} \mid B \preceq B^* \})$ %
            \Else
                \State \Return $\mathrm{\AcceptMerkleProofWrt}(\{ b^*.\mathrm{auxinnov} \} \cup \{ B.\mathrm{auxinnov} \mid B \preceq B^* \})$
            \EndIf
        \EndFunction
    \end{algorithmic}
\end{algorithm}

Pseudocode of SPV is given by Algorithm \ref{algo:pseudocode-light-client},
which also details how \blockslc and \blocksbft are composed
to produce the required light client auxiliary information.
    The function
    $\supersanitize(.)$ takes a sequence of transactions, determines for each transaction its validity with respect to its prefix,
    and outputs the sequence of valid transactions.
    The function
    $\merkleize(.)$ takes a list %
    and constructs a
    vector commitment that also
    allows efficient data availability checks.
    The commitment is part of the block header and
    as such downloaded by light clients,
    while the list itself is only kept by full nodes.
    The function
    $\AcceptMerkleProofWrt(.)$ takes a set of commitments
    with respect to which the light client will accept
    responses provided by full nodes to its queries.
    The relation `$\preceq$' means `is in the prefix of'.
    Following the ledger extraction (Figure~\ref{fig:ledger-extraction}),
    ledgers with respect to the tips $B$ and $b$ of the BFT
    and LC chain, respectively, are defined recursively:
    \begin{IEEEeqnarray}{rClCrCl}
        \LOGlc{}{}(b_{\mathrm{genesis}}) &=& \emptyset
        &\quad&
        \LOGlc{}{}(b) &=& \supersanitize( \LOGlc{}{}(b.\mathrm{prev}) \| b.\Txs{}{})   \IEEEeqnarraynumspace\\
        \LOGfin{}{}(B_{\mathrm{genesis}}) &=& \emptyset
        &\quad&
        \LOGfin{}{}(B) &=& \supersanitize( \LOGfin{}{}(B.\mathrm{prev}) \| \LOGlc{}{}(B.b))   \IEEEeqnarraynumspace\\
        &&
        &\quad&
        \LOGda{}{}(B,b) &=& \supersanitize( \LOGfin{}{}(B) \| \LOGlc{}{}(b))
    \end{IEEEeqnarray}

Finally, we prove the security of the construction,
namely that a light client does not suffer from any safety violations
other than those a full client in its stead
would have also suffered from.
If the client follows $\LOGfin{}{}$,
it is obvious that this is the case.
We proceed to show it for $\LOGda{}{}$.

\begin{theorem}
\label{thm:spv-safety}
If a light client following $\LOGda{}{}$
accepts a transaction $\tx$ as valid at time $t$,
then a full client in its stead would have done the same.
\end{theorem}
Proof is given in Appendix~\ref{sec:appendix-lightclient}.
Note that a trivial SPV procedure that does not accept any transaction satisfies the above security criterion. %
Hence, an SPV procedure that satisfies Theorem~\ref{thm:spv-safety} is only useful if it can also `often' verify the validity of transactions.
The SPV procedure outlined by Algorithm \ref{algo:pseudocode-light-client} can quickly confirm the validity of transactions \emph{under good network conditions}:

\begin{theorem}
\label{thm:spv-liveness}
When $\LOGlc{}{}$ is secure, a light client following $\LOGda{}{}$ can confirm the validity of any transaction as quickly as a full client in the same situation.
\end{theorem}

Proof
is given in Appendix~\ref{sec:appendix-lightclient}.
Recall from
Section \ref{sec:protocols}, $\LOGlc{}{}$ is always secure under
\textbf{P2} (in which case $\LOGda{}{}$ is also live) and it regains its security after $\max\{\GST,\GOT\}$ under \textbf{P1} (in which case $\LOGfin{}{}$ also becomes live).
Hence, whenever $\LOGda{}{}$ is secure and thus $\LOGlc{}{}$ is secure,
or $\LOGfin{}{}$ is live,
by Theorem~\ref{thm:spv-liveness}, any transaction verifiable by a full client is verifiable by a light client.

%% file: appendix_accountability_hotstuff.tex
\section{Accountable Safety for HotStuff}
\label{sec:appendix-accountability-hotstuff}

First, we establish slashing conditions for HotStuff:
\begin{definition}[HotStuff slashing conditions]
Recall that in HotStuff votes contain besides a block
and the view number also a type
(one of: $\text{\textsc{prepare}}$, $\text{\textsc{pre-commit}}$, $\text{\textsc{commit}}$).
A validator's stake is slashed if:
\begin{enumerate}
    \item
        The validator votes for the same type
        and the same view more than once.
        
    \item
        The validator votes for conflicting
        blocks $B_1, B_2$ in two views $v_1 < v_2$
        first $\text{\textsc{commit}}$ in view $v_1$
        and later $\text{\textsc{prepare}}$ in view $v_2$
        (unless there has been a $\frac{2}{3}$-majority
        for $\text{\textsc{prepare}}$
        for an earlier conflicting
        block $B$ in view $v_1 < v < v_2$).

\end{enumerate}
\end{definition}

Again,
$\alpha$-accountable safety has two complementary aspects
(analogous to Lemmas~\ref{thm:accountability-streamlet-attack} and \ref{thm:accountability-streamlet-defense}), which we prove subsequently (for $\alpha=\frac{1}{3}$):
\begin{lemma}
\label{thm:accountability-hotstuff-attack}
In the case of a safety violation in HotStuff,
we can pinpoint $\alpha \cdot n$ validators that must have violated
a slashing condition.
\end{lemma}
\begin{lemma}
\label{thm:accountability-hotstuff-defense}
Honest validators never violate a slashing condition in HotStuff.
\end{lemma}

The proof of Lemma~\ref{thm:accountability-hotstuff-attack}
proceeds along the safety argument of HotStuff.
\begin{proof}
Suppose there is a safety violation.
For this, it is necessary that
there are two quorum certificates (\ie, a collection
of $\geq 2/3$ votes -- think of Streamlet notarizations)
of type $\text{\textsc{commit}}$ for conflicting blocks $B_1, B_2$.

If the certificates are of the same view, $v_1 = v_2$, then
$\geq 1/3$ of validators have voted twice $\text{\textsc{commit}}$
in the same view,
violating the first slashing condition.

So we can assume that the two quorum certificates come from different
views $v_1 \neq v_2$.
Without loss of generality,
assume that the $\text{\textsc{commit}}$ quorum
certificate for $B_1$ comes from a smaller view than that of $B_2$,
$v_1 < v_2$.

Now denote by $v$ the lowest view higher than $v_1$ for which there is
a valid quorum certificate of type $\text{\textsc{prepare}}$
and the referenced block $B$ conflicts with $B_1$.
Such $v$ exists, because at least $v_2$ satisfies the above requirements
(to obtain a $\text{\textsc{commit}}$ quorum certificate for $v_2$,
a $\text{\textsc{prepare}}$ quorum certificate must have been issued before,
otherwise honest validators will not contribute their (required!) shares in
producing the $\text{\textsc{commit}}$ in $v_2$).

Now, $\geq 2/3$ of validators have contributed to the quorum certificates
of type $\text{\textsc{commit}}$ in view $v_1$
and of type $\text{\textsc{prepare}}$ in view $v$.
So, at least $\geq 1/3$ of validators have first voted
$\text{\textsc{commit}}$ in view $v_1$
and later voted $\text{\textsc{prepare}}$ in view $v$,
violating the second slashing condition.

Note that an honest node would not have violated the second slashing
condition in thise case, for the following reason.
Consider a validator $i$ from the intersection of the quorums.
During view $v_1$, $i$ locked on a $\text{\textsc{pre-commit}}$ of $B_1$.
By minimality of $v$, that lock cannot have been changed, as for a change of
the lock another $\text{\textsc{pre-commit}}$ and for that
another $\text{\textsc{prepare}}$ quorum certificate would have been required
(otherwise honest validators do not contribute their votes, which are required
to reach quorum), which contradicts minimality of $v$.

Thus, in view $v$, the proposed block would not have been considered
safe (`\textsc{safeNode}'), as
(a) it is not consistent with the lock on $\text{\textsc{pre-commit}}$ of $B_1$, and
(b) the liveness exception of the safety rule is also not satisfied, because if
it was, \ie, if the justification for the proposal in $v$ is a $\text{\textsc{prepare}}$
quorum certificate that is of view higher than $v_1$ (but lower than $v$,
otherwise it is invalid),
then that contradicts the minimality of $v$.
Thus, in view $v$, the validator $i$ would not have voted $\text{\textsc{prepare}}$
for the proposal that conflicts with $B_1$.
\end{proof}

The proof of Lemma~\ref{thm:accountability-hotstuff-defense}
follows by inspecting the HotStuff pseudocode \cite{yin2018hotstuff}.
\begin{proof}
        An honest validator does not violate the first
        slashing condition
        because it only votes once per type and view.
        An honest validator does not violate the second
        slashing condition
        because of the argument given at the end of the proof
        of Lemma~\ref{thm:accountability-hotstuff-defense}.
\end{proof}

From Lemmas~\ref{thm:accountability-hotstuff-attack} and \ref{thm:accountability-hotstuff-defense}
readily follows:
\begin{theorem}
HotStuff provides $\frac{1}{3}$-accountable safety.
\end{theorem}

%% file: appendix_lightclient.tex
\section{Security Proof for Light Clients and SPV}
\label{sec:appendix-lightclient}

\tikzset{blockchain/.style={
        x=0.3cm,
        y=0.6cm,
        node distance=0.5cm,
        block/.style = {
            minimum width=0.25cm,
            minimum height=0.25cm,
            draw,
            shade,
            top color=white,
            bottom color=black!10,
        },
        link/.style = {
        },
        refsnapshot/.style = {
            {Circle[length=3pt]}-latex,
            myParula01Blue,
            bend right=30,
            shorten <=-1.5pt,
            thick,
        },
        refLOGfin/.style = {
            {Circle[length=3pt]}-latex,
            myParula02Orange,
            bend left=10,
            shorten <=-1.5pt,
            thick,
        },
        refsnapshotLIGHT/.style = {
            refsnapshot,
            myParula01Blue!20,
            thin,
        },
        refLOGfinLIGHT/.style = {
            refLOGfin,
            myParula02Orange!20,
            thin,
        },
    }
}

\subsection{Proof of Theorem \ref{thm:spv-safety}}

\begin{proof}
Assume that a light client following $\LOGda{}{}$ accepts a transaction $\tx$ as valid at time $t$.
Let $b^*$ denote the \blocklc at the tip of the LC chain
as viewed by the light client at time $t$
and $B$ denote the \blockbft referenced (\tikz{ \draw [blockchain,refLOGfin,bend left=20] (0,0) to (2,0); }) by $b^*$.
Similarly, let $B^*$ denote the tip of the BFT chain as viewed by the light client at time $t$.
By line 10 of Algorithm \ref{algo:pseudocode-light-client}, the light client querying $\tx$ follows either the metadata of the \blocksbft in the prefix of $B^*$ or the metadata in $b^*$, only if all of the \blocksbft in the path from $B$ to $B^*$ reference (\tikz{ \draw [blockchain,refsnapshot,bend right=20] (0,0) to (-2,0); }) snapshots that are prefixes of $\LOGlc{}{}(b^*)$.
Then, the transactions preceding $\tx$ within $\LOGda{}{}$ at time $t$  come from either $\LOGlc{}{}(b^*)$ or $\LOGfin{}{}(B)$, with respect to which $\tx$ is valid.
(See the procedure for composing \blockslc in Algorithm \ref{algo:pseudocode-light-client}.)
Consequently, a full client
in the same situation as the light client
would also see $\tx$ as a valid transaction in $\LOGda{}{}$ at time $t$.
\end{proof}

\subsection{Proof of Theorem \ref{thm:spv-liveness}}

\begin{proof}
Assume that a light client following $\LOGda{}{}$ queries a transaction $\tx$ and let $b^*$ denote the tip of the LC chain.
Note that when $\LOGlc{}{}$ is secure, final \blocksbft cannot reference snapshots of $\LOGlc{}{}$ that conflict with each other. 
Then, the \blocksbft on the path from the \blockbft referenced (\tikz{ \draw [blockchain,refLOGfin,bend left=20] (0,0) to (2,0); }) by $b^*$ to the tip of the BFT chain seen by the light client all reference (\tikz{ \draw [blockchain,refsnapshot,bend right=20] (0,0) to (-2,0); }) snapshots that do not conflict with $\LOGlc{}{}(b^*)$.
Hence, SPV can use the metadata in $b^*$ (see lines 10 and 13 of Algorithm \ref{algo:pseudocode-light-client}) through which the light client can verify the validity of $\tx$ as quickly as a full client in the same situation.
\end{proof}